\documentclass[10pt]{article}
\usepackage{latexsym}
\usepackage{amssymb}
\usepackage{amsmath}
\usepackage{amscd}
\usepackage{amsthm}
\usepackage[left=2cm,top=2.5cm,right=2.5cm,bottom=1.5cm]{geometry}
\usepackage[dvips]{graphicx}
\usepackage{hyperref}
\begin{document}
\begin{center}
\large{\bf{Accelerating dark energy models with anisotropic fluid in Bianchi type-$VI_{0}$ space-time}} \\
\vspace{10mm}
\normalsize{Anirudh Pradhan\\
\vspace{5mm}
\normalsize{Department of Mathematics, Hindu Post-graduate College, Zamania-232 331, 
Ghazipur, India; \\
\vspace{2mm}
{\it pradhan@iucaa.ernet.in; pradhan.anirudh@gmail.com}}} \\
\end{center}
\vspace{10mm}
\begin{abstract} 
Motivated by the increasing evidence for the need of a geometry that resembles Bianchi morphology to explain 
the observed anisotropy in the WMAP data, we have discussed some features of the Bianchi type-$VI_{0}$ universes 
in the presence of a fluid that wields an anisotropic equation of state (EoS) parameter in general relativity. 
We present two accelerating dark energy (DE) models with an anisotropic fluid in Bianchi type-$VI_{0}$ space-time. 
To prevail the deterministic solution we choose the scale factor $a(t) = \sqrt{t^{n}e^{t}}$, which yields a 
time-dependent deceleration parameter (DP), representing a class of models which generate a transition of the 
universe from the early decelerating phase to the recent accelerating phase. Under the suitable condition, the 
anisotropic models approach to isotropic scenario. The EoS for dark energy $\omega$ is found to be time-dependent 
and its existing range for derived models is in good agreement with the recent observations of SNe Ia data 
(Knop et al. in Astrophys. J. 598:102, 2003), SNe Ia data with CMBR anisotropy and galaxy clustering statistics 
(Tegmark et al. in Astrophys. J. 606:702, 2004) and latest combination of cosmological datasets coming from CMB 
anisotropies, luminosity distances of high redshift type Ia supernovae and galaxy clustering (Hinshaw et al. in 
Astrophys. J. Suppl. 180:225, 2009, Komatsu et al. in Astrophys. J. Suppl. 180:330, 2009). For different values 
of $n$, we can generate a class of physically viable DE models.The cosmological constant $\Lambda$ is found to be 
a positive decreasing function of time and it approaches to a small positive value at late time (i.e. the present epoch) 
which is corroborated by results from recent type Ia supernovae observations. We also observe that our solutions are 
stable. The physical and geometric aspects of both the models are also discussed in detail. 
\end{abstract}
\smallskip
{\bf Keywords} Cosmological models $.$ Dark energy $.$ Variable EoS parameter \\
\smallskip
{\bf PACS number:} 98.80.Es, 98.80-k, 95.36.+x 
\section{INTRODUCTION} 
Recent cosmological observations obtained by SNe Ia (Riess et al. 1998; Perlmutter et al. 1999)  
suggested that the expansion of the universe is accelerating. Recent observations of SNe Ia of high confidence 
level (Tonry et al. 2003; Riess et al. 2004; Clocchiatti et al. 2006) have further confirmed this. In addition, 
measurements of the cosmic microwave background (CMB) anisotropies (Bennett et al. 2003; de Bernardis et al. 2000, 
Hanany et al. 2000), large scale structure (LSS) (Tegmark et al. 2004a,b; Spergel et al. 2003), the Sloan digital sky 
survey (SDSS) (Seljak et al 2005; Adelman-McCarthy et al. 2006), the Wilkinson microwave anisotropy probe (WMAP) 
(Perlmutter et al. 2003) and the Chandra x-ray observatory (Allen et al. 2004) strongly indicate 
that our universe is dominated by a component with negative pressure, dubbed as dark energy, which constitutes with 
$\simeq 3/4$ of the critical density. The cosmic acceleration is realized with negative pressure and positive energy 
density that violate the strong energy condition. This violation gives a reverse gravitational effect. Due to this effect, 
the Universe gets a jerk and the transition from the earlier deceleration phase to the recent acceleration phase take place 
(Caldwell et al. 2006). A recent survey of more than $200,000$ galaxies appears to confirm the existence 
of dark energy, although the exact physics behind it remains unknown ( Rincon 2011). \\

During the last two decades cosmology is speedily becoming an experimental involvement of physics. The theoretical models 
can be tested, and new and more accurate data in the near future will restrict our conceptions of the Universe to within 
few percent accuracy. The simplest candidate for the dark energy is the cosmological constant (Overduin and Cooperstock 
1998; Sahni and Starobinsky 2000; Komatsu et al. 2009; Kachru et al. 2003) which suffers from conceptual problems such 
as fine-tuning and coincidence problems (Weinberg 1989). Other scenarios include, Quintessence (Wetterich 1988; Ratra 
and Peebles 1988), Chameleon (Khoury and Weltman 2004), K-essence (Chiba et al. 2000; Armendariz-Picon et al. 2000), 
which is based on earlier work of K-inflation (Armendariz-Picon et al. 1999), modified gravity (Capozziello 2002; 
Caroll et al. 2004; Nojiri and Odintsov 2003, 2004; Abdalla et al. 2005; El-Nabulsi 20011a), Tachyon (Padmanabhan 2002) 
arising in string theory (Sen 2002), Quintessential inflation (Peebles and Vilenkin 1999), Chaplygin gas as well as 
generalized Chaplygin gas (Srivastava 2005; Bertolami et al. 2004; Bento et al. 2002; Bilic et al. 2002; Avelino et al. 2003), 
cosmological nuclear energy (Gupta \& Pradhan 2010). Recently, El-Nabulsi (2011b), Feng and Yang (2011), Biesiada et al. 
(2011), Singh and Chaubey (2012), Amirhashchi et al. (2011a) and Pradhan \& Amirhashchi (2011a) have studied DE models 
in different context. In spite of these attempts, still cosmic acceleration is a challenge to modern cosmology and modern 
astrophysics. \\

In general relativity, the evolution of the expansion rate is parameterized by the cosmological equation 
of state (the relationship between temperature, pressure, and combined matter, energy, and vacuum energy density for 
any region of space). Measuring the equation of state for dark energy is one of the biggest efforts in observational 
cosmology today. The DE model has been characterized in a conventional manner by the equation of state (EoS) parameter 
$\omega(t) = \frac{p}{\rho}$ which is not necessarily constant, where $\rho$ is the energy density and $p$ is the fluid 
pressure (Carroll and Hoffman 2003). The present data seem to slightly favour an evolving dark energy with EoS $\omega < -1$ 
around the present epoch and $\omega > -1$ in the near past. Obviously, $\omega$ cannot cross $-1$ for quintessence or 
phantom alone. Some efforts have been made to build a dark energy model whose EoS can cross the phantom divide. The simplest 
DE candidate is the vacuum energy ($\omega = - 1$), which is mathematically equivalent to the cosmological constant 
($\Lambda$). The other conventional alternatives, which can be described by minimally coupled scalar fields, are 
quintessence ($\omega > - 1$) (Steinhardt and Wesley 2009), phantom energy $(\omega < - 1$ (Caldwell 2002) and quintom (that 
can across from phantom region to quintessence region as evolved) and have time dependent EoS parameter. Some other limits 
obtained from observational results coming from SNe Ia data (Knop et al. 2003) and combination of SNe Ia data with CMBR 
anisotropy and galaxy clustering statistics (Tegmark et al. 2004a,b) are $-1.67 < \omega < -0.62$ and $-1.33 < \omega < - 0.79$, 
respectively. The latest results in 2009, obtained after a combination of cosmological datasets coming from CMB anisotropies, 
luminosity distances of high redshift type Ia supernovae and galaxy clustering, constrain the dark energy EoS to 
$-1.44 < \omega < -0.92$ at $68\%$ confidence level (Hinshaw et al. 2009; Komatsu et al. 2009). However, it is not at 
all obligatory to use a constant value of $\omega$. Due to lack of the observational evidence in making a distinction 
between constant and variable $\omega$, usually the equation of state parameter is considered as a constant 
(Kujat et al. 2002; Bartelmann et al. 2005; Yadav 2011) with phase wise value $-1, 0, - \frac{1}{3}$ and $ + 1$ for 
vacuum fluid, dust fluid, radiation and stiff dominated universe, respectively. But in general, $\omega$ is a function 
of time or redshift $z$ or scale factor $a$ as well (Ratra and Peebles 1988; Jimenez 2003; Das et al. 2005). In earlier 
various form of time dependent $\omega$ have been used for variable $\Lambda$ models by Mukhopadhyay et al. (2008). 
Recently, dark energy models with variable EoS parameter have been studied by Ray et al. (2010), Akarsu and Kilinc 
(2010a,b), Yadav et al. (2010), Yadav and Yadav (2010), Pradhan and Amirhashchi (2011a), Pradhan et al. (2011), 
Amirhashchi et al. (2011a,b) and Saha \& Yadav (2012). In well-known reviews on modified gravity (Nojiri and 
Odintsov 2007, 2011), it is clearly indicated that any modified gravity may be represented as effective fluid with 
time dependent $\omega$. The dark energy universe EoS with inhomogeneous, Hubble parameter dependent term is considered 
by Nojiri and Odintsov (2005). Later on, Nojiri and Odintsov (2006) have also presented the late-time cosmological 
consequences of dark energy with time-dependent periodic EoS in oscillating universe. \\

Today there is considerable evidence, which suggests that the universe may be isotropic and homogeneous. After 
discovery of CMB radiation, cosmology became a precision science. The CMB radiation is also considered to be a 
major experimental evidence on which the most commonly accepted theory about the origin of universe, i.e. ``Big-Bang'' 
cosmology, is based. Statistical Isotropy (SI) is usually assumed in almost all CMB studies. But, now, there exist 
many indications which suggest that CMB may violate this assumption. Apart from the CMB there are some other indications 
of violation of SI which suggest the existence of a preferred direction in the universe. These indications include 
distributions of polarizations from radio galaxies (Birch 1982; Jain \& Ralston 1999; Jain et al. 2004) and statistics 
of optical polarizations from quasars (Hutsem$\acute{e}$kers 1998; Hutsem$\acute{e}$kers \& Lamy 2001; Jain et al. 2004; 
Ralston \& Jain 2004). Polarization of electromagnetic waves coming from distant Radio Galaxies and Quasars measured 
at radio and optical frequencies respectively are not consistent with the assumptions of SI, rather radio polarizations 
are organized coherently over the dome of the sky and optical polarizations are aligned in a preferential direction on 
very large scales, violating the assumed isotropy of the universe. These study confirmed strong significance of anisotropy 
and also claimed that the statistics are not consistent with isotropy at $99.9\%$ confidence level. It has also been observed 
that the quadrupole and the octopole have almost all their power perpendicular to a common axis in space pointing towards 
Virgo cluster (Tegmark et al. 2003; de Oliveira-Costa 2004). The dipole, which is commonly attributed to our motion 
relative to the CMB rest frame, also aligns in the same direction as quadrupole and octopole which is not expected under the 
condition of statistical isotropy. Another indication of anisotropy in CMB data is the presence of a cold spot with 
improbably low temperature. It was found by Cruz et al. (2005) by using Spherical Mexican Hat Wavelet analysis on WMAP data.
Several authors have also searched for anisotropy using the supernova data set. Jain et al. (2007) found violation of 
isotropy in this data. Subsequently, there have been a large number of studies (Bielewicz et al. 2004; Eriksen et al. 2004; 
Katz \& Weeks 2004; Bielewicz et al. 2005; Prunet et al. 2005; Bernui et al. 2006; de Oliveira-Costa \& Tegmark 2006; 
Freeman et al. 2006; Bernui et al. 2007; Land \& Magueijo 2007) which claim the CMB is not consistent with isotropy. The 
possible violation of SI in CMB has lead to many theoretical studies. Several physical explanations for the observed 
anisotropy  have been put forward (Cline et al. 2003; Contaldi et al. 2003; Kesden et al. 2003; 
Armend$\acute{a}$riz-Pic$\acute{o}$n 2004; Berera et al. 2004; Gordon et al. 2005; Abramo et al. 2006; Campanelli et al. 
2007; Rodrigues 2008). Land \& Magueijo (2005) found evidence that the detected anisotropy has 
positive mirror parity. The generation and evolution of primordial perturbations in an anisotropic universe has also been 
studied (Armend$\acute{a}$riz-Picon 2006; Battye \& Moss 2006; Pereira et al. 2007) along with the possibility of
anisotropic inflation (Hunt \& Sarkar 2004; Buniy et al. 2006; Donoghue et al. 2009). \\

The possible violation of global isotropy in the CMB has been a subject of intense research after the publication of 
WMAP data. The possible alignment of axes corresponding to several diverse data sets in the direction of the Virgo 
cluster makes this extremely interesting. In recent years, there have been a large number of studies which claim the 
CMB temperature fluctuations are not consistent with statistical isotropy and thus questioning the cosmological principle. 
The CMB is considered to be a major experimental evidence which supports the current/present models of the observed universe 
and from this CMB observations several people found significant of anisotropic scenario. Based on these studies one may not 
preclude the possibility that our universe is anisotropic. \\

There is a general agreement among cosmologists that cosmic microwave's background anisotropy in the small angle 
scale holds the key to the formation of the discrete structure. The theoretical argument (Misner 1968) and the modern 
experimental data support the existence of an anisotropic phase, which turns into an isotropic one. The anisotropy 
of the DE within the framework of Bianchi type space-times is found to be useful in generating arbitrary ellipsoidality 
of the Universe, and to fine tune the observed CMBR anisotropies. Koivisto and Mota (2008a, 2008b) have investigated 
cosmological models with anisotropic EoS and have also shown that the present SN Ia data allows large anisotropy. 
Recently, Akarsu and Kilinc (2010c) have described some features of the Bianchi type-I universes in presence of fluid 
that wields an anisotropic EoS. Hence, for a realistic cosmological model one should consider spatially homogeneous 
and anisotropic space-times and then show whether they can evolve to the observed amount of homogeneity and isotropy. 
The only spatially homogeneous but anisotropic models other than Bianchi type models are the Kantowski-Sachs locally 
symmetric family. See Ellis \& van Elst (1999) for generalized, particularly anisotropic, cosmological models and 
Ellis (2006) for a concise review on Bianchi type models. The motivation for this investigation comes from the hints 
of statistical anisotropy of our Universe that several observations seem to suggest. \\  
 
Bianchi type-$VI_{0}$ (B-$VI_{0}$) space-time in connection with massive strings is studied by Pradhan and Bali (2008) 
and Bali et al. (2008). Belinchon (2009) studied several cosmological models with B-$VI_{0}$ \& III symmetries 
under the self similar approach. Given the growing interest of cosmologists, here, we propose to study the evolution of the 
Universe within the framework of a B-$VI_0$ space-time. Recently, Amirhashchi et al. (2011c) and Pradhan et al. (2012) 
presented dark energy models in an anisotropic B-$VI_0$ space-time by considering constant and variable deceleration 
parameter respectively. In this paper, we have investigated two new B-$VI_{0}$ DE models with variable $\omega$ by 
assuming different scale factors in such a way that they provide time dependent deceleration parameter in presence of 
anisotropic fluid. The out line of the paper is as follows: In Sect. $2$, the metric and the field equations are described. 
Section $3$ deals with the solutions of the field equations. Section $4$  deals with physical and geometric behaviour of 
the model. Section $5$ deals with the stability of the corresponding solutions. In Section $6$, we describe an other dark 
energy model and its physical aspects. In Sect. $7$, we again examine the stability of corresponding solutions for second 
DE model. Finally, conclusions are summarized in the last Sect. $8$. 
\section{THE METRIC AND FIELD EQUATIONS}
We consider totally anisotropic Bianchi type-$VI_{0}$ line element, given by
\begin{equation}
\label{eq1}
ds^{2} = - dt^{2}+ A^{2}dx^{2} + B^{2}e^{2x} dy^{2} + C^{2}e^{-2x} dz^{2},
\end{equation}
where the metric potentials $A$, $B$ and $C$ are functions of $t$ alone. This ensures that the model is
spatially homogeneous. \\

The simplest generalization of EoS parameter of perfect fluid may be to determine the EoS 
parameter separately on each spatial axis by preserving the diagonal form of the energy momentum tensor in a 
consistence way with the considered metric. Therefore, the energy momentum tensor of fluid can be written, most generally, 
in anisotropic diagonal form as follows:
\begin{equation}
\label{eq2}
T^{j}_{i} = diag [T^{0}_{0}, T^{1}_{1}, T^{2}_{2}, T^{3}_{3}]
\end{equation}
Allowing for anisotropy in the pressure of the fluid, and thus in its EoS parameter, gives rise to new possibilities 
for the evolution of the energy source. To see this, we first parametrize the energy momentum tensor given in (\ref{eq2}) 
as follows:
\[
T^{j}_{i} = {\rm diag}[\rho, - p_{x}, - p_{y}, - p_{z}] 
\]
\[
= {\rm diag}[1, - \omega_{x}, - \omega_{y}, - \omega_{z}]\rho
\]
\begin{equation}
\label{eq3} =  {\rm diag}[1, - \omega, - (\omega + \delta), - (\omega + \gamma)]\rho.
\end{equation}
Here $\rho$ is the proper energy density, $p_{x}, p_{y}$ and $p_{z}$ are the pressures, 
and $\omega_{x}, \omega_{y}$ and $\omega_{z}$ are the directional EoS parameters along the $x$, $y$ and $z$ axes, 
respectively; $\omega$ is the deviation-free EoS parameter of the fluid. The deviation 
from isotropy is parametrized by setting $\omega_{x}= \omega$ and then introducing skewness parameters $\delta$ and $\gamma$ 
which are the deviations from $\omega$, respectively along the $y$ and $z$. $\omega$, $\delta$ and $\gamma$ are not 
necessarily constants and might be function of the cosmic time, $t$.\\

The Einstein's field equations (with gravitational units, $8\pi G = 1$ and $c = 1$) read as
\begin{equation}
\label{eq4} R^{j}_{i} - \frac{1}{2} R g^{j}_{i} = - T^{j}_{i},
\end{equation}
where the symbols have their usual meaning. In a comoving co-ordinate system, Einstein's field equation (\ref{eq4}), 
with (\ref{eq3}) for B-$VI_{0}$ metric (\ref{eq1}) subsequently lead to the following system of equations:
\begin{equation}
\label{eq5} 
\frac{\ddot{B}}{B} + \frac{\ddot{C}}{C} + \frac{\dot{B}\dot{C}}{BC} + \frac{1}{A^{2}} = - \omega\rho,
\end{equation}
\begin{equation}
\label{eq6} 
\frac{\ddot{C}}{C} + \frac{\ddot{A}}{A} + \frac{\dot{C}\dot{A}}{CA} - \frac{1}{A^{2}} = - (\omega + \delta)\rho,  
\end{equation}
\begin{equation}
\label{eq7} 
\frac{\ddot{A}}{A} + \frac{\ddot{B}}{B} + \frac{\dot{A}\dot{B}}{AB} - \frac{1}{A^{2}} = - (\omega + \gamma)\rho,
\end{equation}
\begin{equation}
\label{eq8} 
\frac{\dot{A}\dot{B}}{AB} + \frac{\dot{B}\dot{C}}{BC} + \frac{\dot{C}\dot{A}}{CA} - \frac{1}{A^{2}} = \rho,
\end{equation}
\begin{equation}
\label{eq9} \frac{\dot{C}}{C} - \frac{\dot{B}}{B} = 0.
\end{equation}
Here and in what follows an over dot denotes ordinary differentiation with respect to $t$.\\

The spatial volume for the model (\ref{eq1}) is given by
\begin{equation}
\label{eq10} V^{3} = ABC.
\end{equation}
We define $a = (ABC)^{\frac{1}{3}}$ as the average scale factor so that the Hubble's parameter
is anisotropic and may be defined as
\begin{equation}
\label{eq11} H = \frac{\dot{a}}{a} = \frac{1}{3}\left(\frac{\dot{A}}{A} + \frac{\dot{B}}{B} + \frac{\dot{C}}{C}\right).
\end{equation}
The deceleration parameter $q$, the scalar expansion $\theta$, shear scalar $\sigma^{2}$ and the average anisotropy 
parameter $A_{m}$ are defined by
\begin{equation}
\label{eq12} q = - \frac{a\ddot{a}}{\dot{a}^{2}},
\end{equation}
\begin{equation}
\label{eq13}
\theta = \frac{\dot{A}}{A} + \frac{\dot{B}}{B} + \frac{\dot{C}}{C},
\end{equation}
\begin{equation}
\label{eq14}
\sigma^{2} = \frac{1}{2}\left(\sum^{3}_{i = 1}H^{2}_{i} - \frac{1}{3}\theta^{2}\right),
\end{equation}
\begin{equation}
\label{eq15} A_{m} = \frac{1}{3}\sum_{i = 1}^{3}{\left(\frac{\triangle
H_{i}}{H}\right)^{2}},
\end{equation}
where $\triangle H_{i} = H_{i} - H (i = x, y, z)$.
\section{SOLUTIONS OF THE FIELD EQUATIONS}
Integrating Eq. (\ref{eq9}), we obtain
\begin{equation}
\label{eq16} C = \ell B,
\end{equation}
where $\ell$ is an integrating constant. Now if we put the value of Eq. (\ref{eq16}) in (\ref{eq7}) 
and subtract the result from Eq. (\ref{eq6}), we obtain that the skewness parameters along $y$ and $z$ axes are 
equal, i.e $\delta = \gamma$. \\

Therefore, equations (\ref{eq5})-(\ref{eq9}) are reduced to
\begin{equation}
\label{eq17} 2\frac{\ddot{B}}{B} + \frac{\dot{B}^{2}}{B^{2}} + \frac{1}{A^{2}} = -\omega\rho,
\end{equation}
\begin{equation}
\label{eq18} \frac{\ddot{A}}{A} + \frac{\ddot{B}}{B} + \frac{\dot{A}\dot{B}}{AB} - \frac{1}{A^{2}} = - 
(\omega + \gamma)\rho,
\end{equation}
\begin{equation}
\label{eq19} 2\frac{\dot{A}\dot{B}}{AB} + \frac{\dot{B}^{2}}{B^{2}} - \frac{1}{A^{2}} = \rho.
\end{equation}
The field equations (\ref{eq17})-(\ref{eq19}) are a system of three linearly independent equations with five 
unknown parameters $A$, $B$, $\omega$, $\rho$ and $\gamma$. Two additional constraints relating these parameters 
are required to obtain explicit solutions of the system. \\

In literature it is common to use a constant deceleration parameter (Akarsu and Kilinc 2010a, 2010b; Amirhashchi et al. 
2011c; Pradhan et al. 2011; Kumar and Yadav 2011; Yadav 2011), as it duly gives a power law for metric 
function or corresponding quantity. The motivation to choose such time dependent DP is behind the fact that the universe 
is accelerated expansion at present as observed in recent observations of Type Ia supernova (Riess et al. 1998; 
Perlmutter et al. 1999; Tonry et al. 2003; Riess et al. 2004; Clocchiatti et al. 2006) and CMB anisotropies (Bennett et al. 
2003; de Bernardis et al. 2000; Hanany et al. 2000) and decelerated expansion in the past. Also, the transition redshift 
from deceleration expansion to accelerated expansion is about 0.5. Now for a Universe which was decelerating in past 
and accelerating at the present time, the DP must show signature flipping (see the Refs. Padmanabhan and Roychowdhury 
(2003), Amendola (2003), Riess et al. (2001)). So, in general, the DP is not a constant but time variable. This motivates 
to choose such a scale factor which yields a time-dependent DP. At this juncture it should be stated that some authors 
first choose the scale factors in power law, exponential or in other form and then find out other variables with some 
conditions under these solutions.\\

In this paper, following Saha et al. (2011) and Pradhan \& Amirhashchi (2011b), we take following {\it ansatz} for the 
scale factor, where increase in term of time evolution is
\begin{equation}
\label{eq20} a(t) = \sqrt{t^{n}e^{t}},
\end{equation}
where $n$ is a positive constant. Saha et al. (2011) and Pradhan \& Amirhashchi (2011b) examined the relation (\ref{eq20}) 
in studying two-fluid scenario for dark energy models in an FRW universe and accelerating DE models in Bianchi type-V 
space-times respectively. This {\it ansatz} generalized the one proposed by Amirhashchi et al. (2011b). If we put $n = 0$ 
in Eq. (\ref{eq20}), it is reduced to $a(t) = \sqrt{e^{t}}$ i.e. a exponential law of variation of scale factor. This 
choice of scale factor yields a time-dependent deceleration parameter (see Eq. (\ref{eq30})) such that before DE era, the 
corresponding solution gives inflation and radiation/matter dominance era with subsequent transition from deceleration to 
acceleration. Thus, our choice of scale factor is physically acceptable. \\

It is worth mention here that one can also select many to many {\it ansatz} other than Eq. (\ref{eq20}) which mimic accelerating 
universe but one should also be careful to check the physical acceptability and stability of their corresponding solutions otherwise  
does not prove any relation of such solutions with observable universe. Eq. (\ref{eq20}) yields physically plausible solutions.\\

Secondly, we assume that the expansion ($\theta$) is proportional to shear ($\sigma$). This condition and Eq. 
(\ref{eq16}) lead to 
\begin{equation}
\label{eq21}
 \frac{1}{\sqrt{3}}\left(\frac{\dot{A}}{A} - \frac{\dot{B}}{B}\right) = \alpha_{0}\left(\frac{\dot{A}}{A} 
+ 2\frac{\dot{B}}{B}\right), 
\end{equation}
which yields to
\begin{equation}
\label{eq22}
 \frac{\dot{A}}{A} = m \frac{\dot{B}}{B},
\end{equation}
where $m = \frac{2\alpha_{0}\sqrt{3} + 1}{1 - \alpha_{0}\sqrt{3}}$ and $\alpha_{0}$ are arbitrary constants. Above equation, 
after integration, reduces to
\begin{equation}
\label{eq23}
 A = \beta (B)^{m},
\end{equation}
where $\beta$ is an integrating constant. Here, for simplicity and without any loss of generality, we assume
$\beta = 1$. Hence we have 
\begin{equation}
\label{eq24} A = (B)^{m}.
\end{equation}
Collins et al. (1980) have pointed out that for spatially homogeneous metric, the normal congruence to the 
homogeneous expansion satisfies that the condition $\frac{\sigma}{\theta}$ is constant. \\

Using equations (\ref{eq16}), (\ref{eq20}) and (\ref{eq24}) in (\ref{eq11}), we obtain the expressions for 
metric functions as follows 
\begin{equation}
\label{eq25}
B(t) = \ell_{1}(t^{n}e^{t})^{\frac{3}{2(m + 2)}},
\end{equation}
\begin{equation}
\label{eq26}
C(t) = \ell_{2}(t^{n}e^{t})^{\frac{3}{2(m + 2)}},
\end{equation}
\begin{equation}
\label{eq27}
A(t) =  \ell_{3}(t^{n}e^{t})^{\frac{3m}{2(m + 2)}},
\end{equation}
where, $\ell_{1} = k^{-\frac{1}{(m + 2)}}, ~ \ell_{2} = \ell \ell_{1}, ~ \ell_{3} = \ell_{1}^{m}$ and $k$ is an 
integrating constant. \\

Hence the model (\ref{eq1}) reduces to
\[
ds^{2} = - dt^{2} + \ell_{3}^{2}(t^{n}e^{t})^{\frac{6m}{(m + 2)}} dx^{2} +  \ell_{1}^{2}(t^{n}e^{t})^{\frac{6}
{(m + 2)}} dy^{2}
\]
\begin{equation}
\label{eq28}
+ \ell_{2}^{2}(t^{n}e^{t})^{\frac{6}{(m + 2)}}dz^{2}.
\end{equation}

\begin{figure}[ht]
\centering
\includegraphics[width=10cm,height=10cm,angle=0]{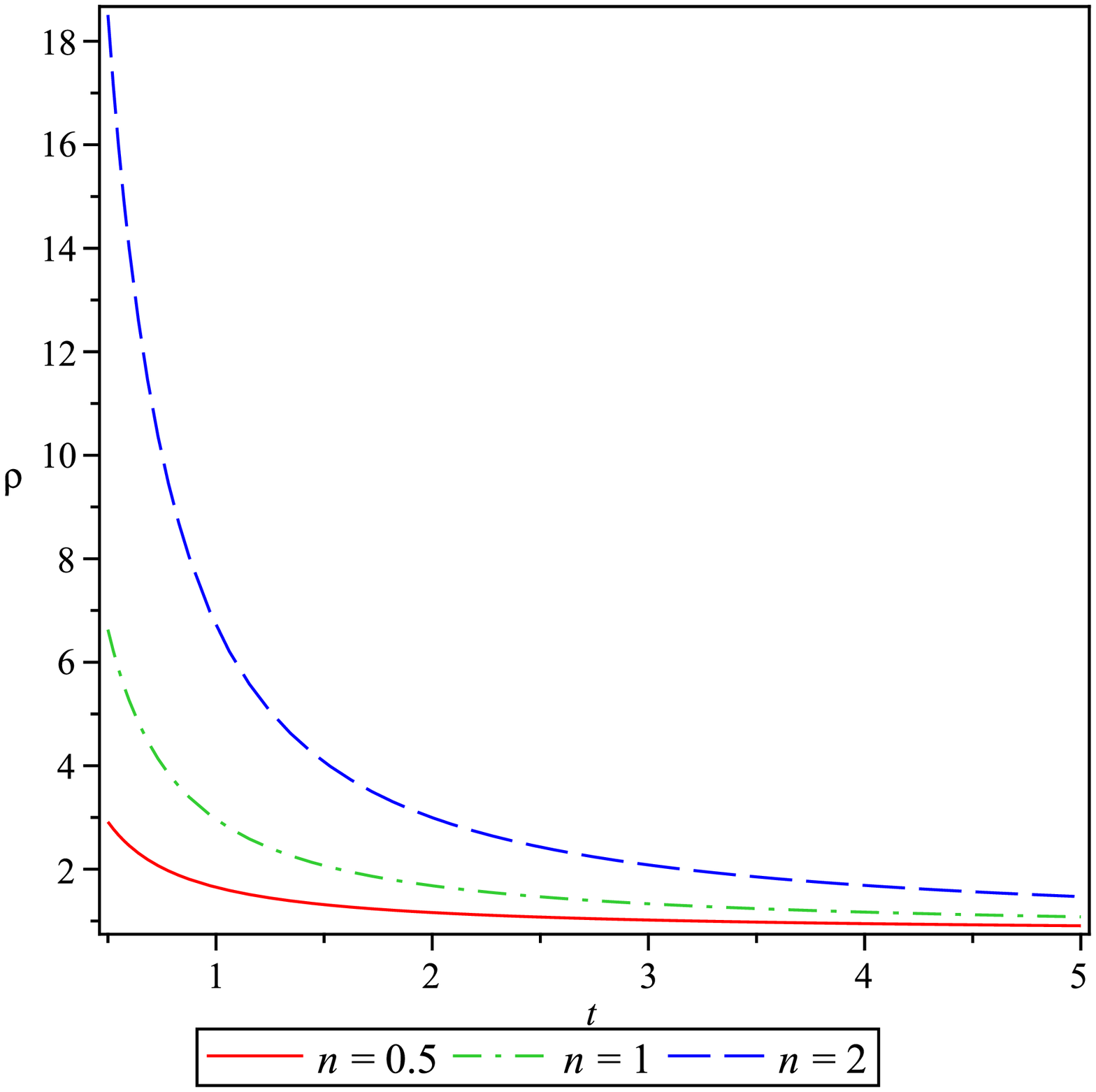} \\
\caption{The plot of energy density$\rho$ versus $t$. 
Here  $\ell_{0} = 0.1$, $m = 1$}.
\end{figure}
\begin{figure}[ht]
\centering
\includegraphics[width=10cm,height=10cm,angle=0]{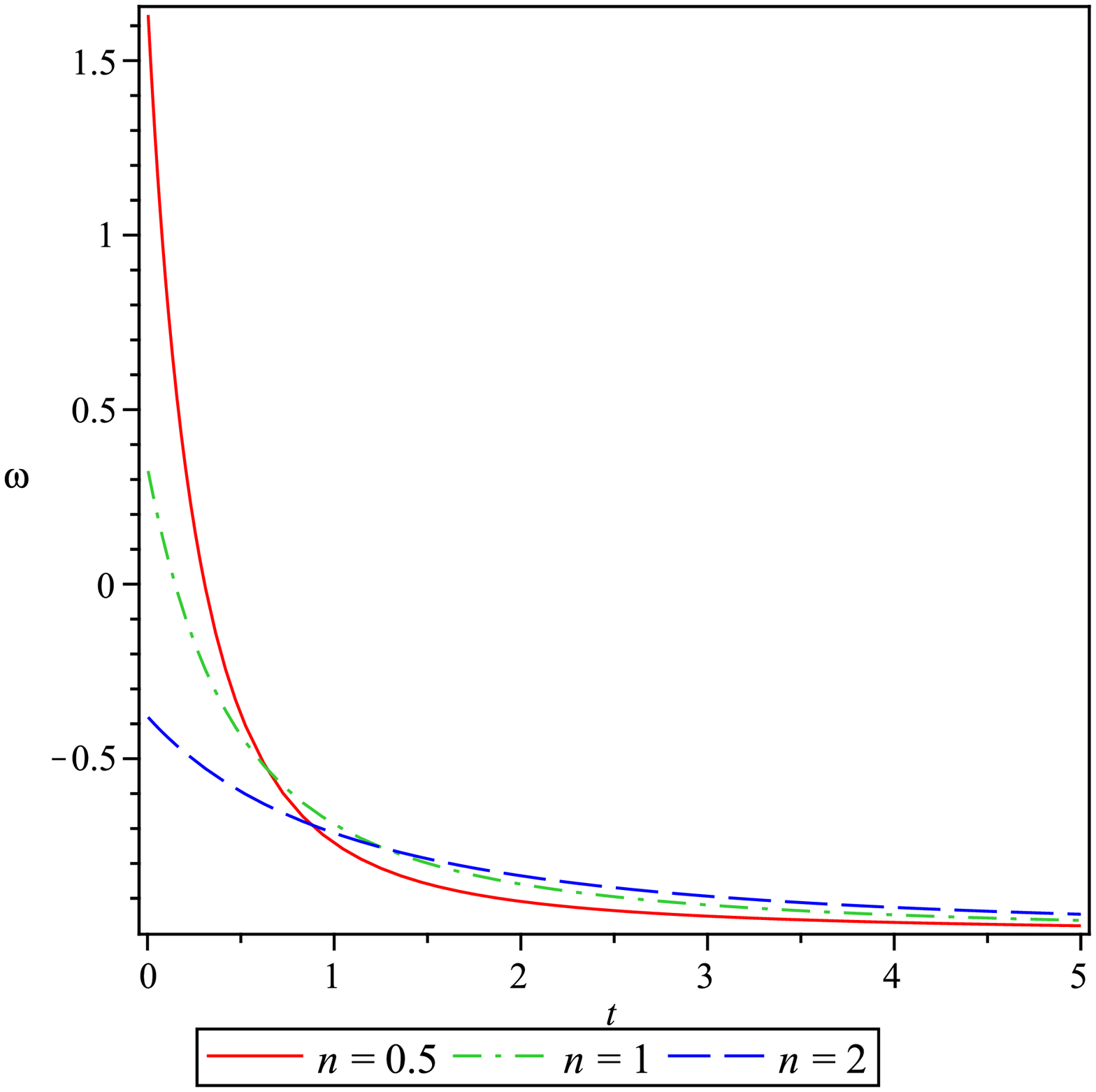} \\
\caption{The plot of EoS parameter $\omega$ versus $t$. 
Here  $\ell_{0} = 0.1$, $m = 1$}.
\end{figure}
\begin{figure}[ht]
\centering
\includegraphics[width=10cm,height=10cm,angle=0]{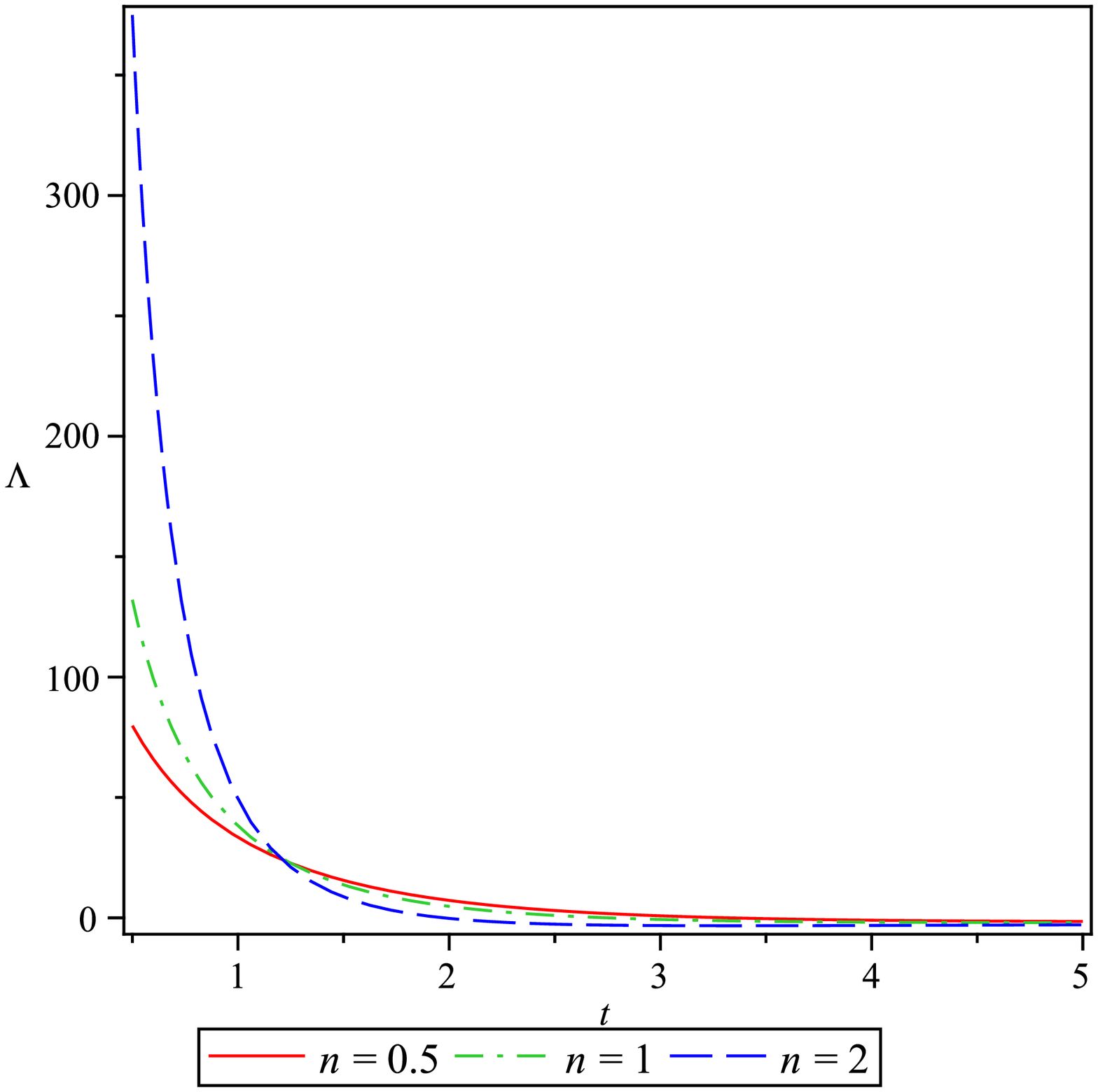} \\
\caption{The plot of cosmological constant $\Lambda$ versus $t$. 
Here $\ell_{0} = 0.1$, $m = 1$}.
\end{figure}
\begin{figure}[ht]
\centering
\includegraphics[width=10cm,height=10cm,angle=0]{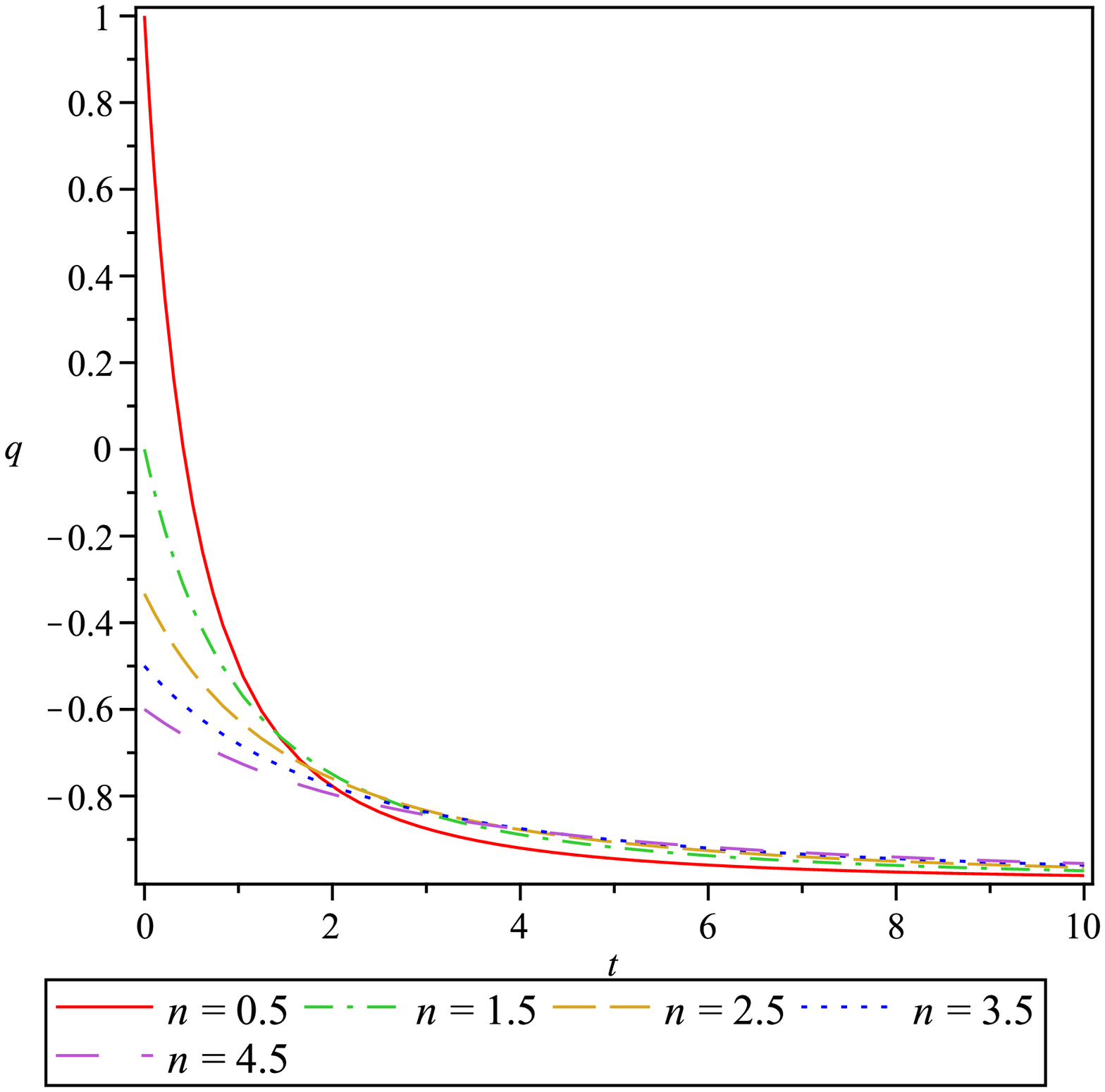} \\
\caption{The plot of deceleration parameter $q$ versus $t$}. 
\end{figure}
\section{PHYSICAL ASPECTS OF THE DARK ENERGY MODEL}
The expressions for the Hubble parameter ($H$), scalar of expansion ($\theta$), shear scalar ($\sigma$), the 
spatial volume ($V$) and the average anisotropy parameter ($A_{m}$) for the model (\ref{eq28}) are given by
\begin{equation}
\label{eq29}
\theta = 3H = \frac{3}{2}\left(1 + \frac{n}{t}\right),
\end{equation}
\begin{equation}
\label{eq30}
q = \frac{2n}{(n + t)^{2}} - 1,
\end{equation}
\begin{equation}
\label{eq31}
\sigma^{2} = \frac{3}{4}\left(\frac{m - 1}{m + 2}\right)^{2}\left(1 + \frac{n}{t}\right)^{2},
\end{equation}
\begin{equation}
\label{eq32}
 V = (t^{n}e^{t})^{\frac{3}{2}},
\end{equation}
\begin{equation}
\label{eq33}
A_{m} =  2\left(\frac{m - 1}{m + 2}\right)^{2}.
\end{equation}
From Eqs. (\ref{eq29})$-$(\ref{eq33}), it is observed that at $t = 0$, the spatial volume vanishes and other parameters 
$\theta$, $\sigma$, $H$ diverge. Hence the model starts with a big bang singularity at $t = 0$. This is a Point Type 
singularity (MacCallum 1971) since directional scale factor $A(t)$, $B(t)$ and $C(t)$ vanish at initial time. Since 
$\frac{\sigma^{2}}{\theta^{2}} \ne 0$ except $m = 1$, hence the model is anisotropic for all values of $m$ except 
for $m \ne 1$. The dynamics of the mean anisotropy parameter depends on the value of $m$. We observe that when $m =1$, 
$A_{m} = 0$ (i.e. the case of isotropy). Thus, the observed isotropy of the model can be achieved in cosmological 
constant region (see, Fig. $2$). \\

The energy density of the fluid can be find by using Eqs. (\ref{eq25}) \& (\ref{eq27}) in (\ref{eq19})
\begin{equation}
\label{eq34}
\rho = \frac{9}{4}\left(\frac{2m + 1}{m + 2}\right)\left(1 + \frac{n}{t}\right)^{2} - 
\ell_{0}(t^{n}e^{t})^{-\frac{3m}{(m + 2)}}.
\end{equation}
where $\ell_{0} = \frac{1}{\ell_{3}^{2}}$. Using Eqs. (\ref{eq25}), (\ref{eq27}) and (\ref{eq34}) in (\ref{eq17}), 
the EoS parameter $\omega$ is obtained as
\begin{equation}
\label{eq35}
\omega = \frac{\frac{27}{4(m + 2)^{2}}\left(1 + \frac{n}{t}\right)^{2} - \frac{3n}{(m + 2)t^{2}} + 
\ell_{0}(t^{n}e^{t})^{-\frac{3m}{(m + 2)}}}{\ell_{0}(t^{n}e^{t})^{-\frac{3m}{(m + 2)}} + 
\frac{9}{4}\left(\frac{2m + 1}{m + 2}\right)\left(1 + \frac{n}{t}\right)^{2}}.
\end{equation}
Using Eqs. (\ref{eq25}), (\ref{eq27}), (\ref{eq34}) and (\ref{eq35}) in (\ref{eq18}), the 
skewness parameters $\delta$ (or $\gamma$) (i.e. deviations from $\omega$ along y-axis and z-axis) 
are computed as
\begin{equation}
\label{eq36}
\delta = \gamma = \frac{\frac{3}{4}\left(\frac{m - 1}{m + 2}\right)\left\{\left(1 + \frac{n}{t}\right)^{2} - \frac{2n}{t^{2}}
\right\} - 2 \ell_{0}(t^{n}e^{t})^{-\frac{3m}{(m + 2)}}}{\ell_{0}(t^{n}e^{t})^{-\frac{3m}{(m + 2)}} -\frac{9}{4}
\left(\frac{2m + 1}{m + 2}\right)\left(1 + \frac{n}{t}\right)^{2}}.
\end{equation}
From equation (\ref{eq35}), it is observed that the equation of state parameter $\omega$ is time dependent, it can be 
function of redshift $z$ or scale factor $a$ as well (as already discussed in previous Section $1$). \\

So, if the present work is compared with experimental results (Knop et al. 2003; Tegmark et al. 2004b; Hinshaw et al. 
2009; Komatsu et al. 2009), then one can conclude that the limit of $\omega$ provided by equation (\ref{eq35}) may 
accommodated with the acceptable range of EoS parameter. Also it is observed that at $t = t_{c}$, $\omega$ vanishes, 
where $t_{c}$ is a critical time given by the following relation
\begin{equation}
\label{eq37}\frac{27}{4(m + 2)^{2}}\left(1 + \frac{n}{t_{c}}\right)^{2} - \frac{3n}{(m + 2)t_{c}^{2}} + 
\ell_{0}(t_{c}^{n}e^{t_{c}})^{-\frac{3m}{(m + 2)}}.
\end{equation}
Thus, for this particular time, our model represents a dusty universe. We also note that the earlier real matter at 
$ t\leq t_{c}$, where $\omega \geq 0$ later on at $t > t_{c}$, where $\omega < 0$ converted to the dark energy 
dominated phase of universe.\\

From Eq. (\ref{eq34}), we note that energy density of the fluid $\rho(t)$ is a decreasing function of time and 
$\rho \geq 0$ when
\begin{equation}
\label{eq38} \left(1 + \frac{n}{t}\right)^{2}(t^{n}e^{t})^{\frac{3m}{(m + 2)}} \geq 
\frac{4\ell_{0}}{9}\left(\frac{m + 2}{2m + 1}\right). 
\end{equation}
Figure $1$ is the plot of energy density of the fluid ($\rho$) versus time in accelerating mode of the universe. 
Here we observe that $\rho$ is a positive decreasing function of time and it approaches to zero as $t \to \infty$. \\

Figure $2$ depicts the variation of EoS parameter ($\omega$) versus cosmic time ($t$) in evolution of the universe, 
as a representative case with appropriate choice of constants of integration and other physical parameters using 
reasonably well known situations (parameters are given in Figure caption). For $m = 1$, we obtain isotropic model 
which is studied here as a representative case. From Figure $2$, we observed that at the initial time there is quintessence 
($\omega > -1$) region and at late time it approaches to the cosmological constant ($\omega = -1$) scenario. This is a 
situation in early universe where quintessence dominated universe (Caldwell 2002) may be playing an important role for 
EoS parameter. Since $\omega$ approaches to $-1$ for sufficiently large time, so its value is consistent with the range 
of all the three observations (Knop et al. 2003; Tegmark et al. 2004b; Hinshaw et al. 2009; Komatsu et al. 2009).\\

In absence of any curvature, matter energy density $\Omega_{m}$ and dark energy $\Omega_{\Lambda}$ are related 
by the equation
\begin{equation}
\label{eq39}
\Omega_{m} + \Omega_{\Lambda} = 1,
\end{equation}
where $\Omega_{m} = \frac{\rho}{3H^{2}}$ and $\Omega_{\Lambda} = \frac{\Lambda}{3H^{2}}$. Thus, equation 
(\ref{eq39}) reduces to
\begin{equation}
\label{eq40}
\frac{\rho}{3H^{2}} + \frac{\Lambda}{3H^{2}} = 1.
\end{equation}
Using Eqs. (\ref{eq29}) and (\ref{eq34}) in (\ref{eq40}), the cosmological constant is obtained as
\begin{equation}
\label{eq41}
\Lambda = -\frac{3}{4}\left(\frac{5m + 1}{m + 2}\right)\left(1 + \frac{n}{t}\right)^{2} + 
\ell_{0}(t^{n}e^{t})^{-\frac{3m}{(m + 2)}}. 
\end{equation}
From Eq. (\ref{eq41}), we observe that $\Lambda$ is a decreasing function of time and it is always positive when
\begin{equation}
\label{eq42} \left(1 + \frac{n}{t}\right)^{2}(t^{n}e^{t})^{\frac{3m}{(m + 2)}} 
< \frac{4\ell_{0}}{3}\left(\frac{m + 2}{5m + 1}\right).
\end{equation}
In general relativity, the Bianchi identities for the Einstein's tensor $G_{ij}$ and the vanishing covariant divergence 
of the energy momentum tensor $T_{ij}$ together with imply that the cosmological term $\Lambda$ is constant. In theories 
with a variable $\Lambda$-term, one either introduces new terms (involving scalar fields, for instance) in to the left hand 
side of the Einstein's field equations to cancel the non-zero divergence of $\Lambda g_{ij}$ (Bergmann 1968; Wagoner 1970) 
or interprets $\Lambda$ as a matter source and moves it to the right hand side of the field equations (Zeldovich 1968), in 
which case energy momentum conservation is understood to mean $T^{*ij}_{;j} = 0$, where $T^{*}_{ij} = T_{ij} - 
(\Lambda/8\pi G)g_{ij}$. It is here that the first assumption that leads to the cosmological constant problem is made. It 
is that the vacuum has a non-zero energy density. If such a vacuum energy density exists, Lorentz invariance requires that 
it has the form $\langle T_{\mu \nu} \rangle = - \langle \rho \rangle g_{\mu\nu}$. This allows to define an effective 
cosmological constant and a total effective vacuum energy density $\Lambda_{eff} = \Lambda + 8\pi G \langle\rho\rangle$ or 
$\rho_{vac} = \langle \rho \rangle + \Lambda/8\pi G$. Note at this point that only the effective cosmological constant, 
$\Lambda_{eff}$, is observable, not $\Lambda$, so the latter quantity may be referred to as a `bare'. The two approaches 
are of course equivalent for a given theory (Vishwakarma 2000). For detail discussions, the readers are advised to see 
the references (Carroll et al. 1992; Abdussattar and Vishwakarma 1996; Peebles and Ratra 2003; Sahni and Starobinsky 2000; 
Padmanabhan 2003, 2008).\\

Figure $3$ is the plot of cosmological constant $\Lambda$ versus time $t$. We observe that cosmological parameter 
is decreasing function of time and it approaches a small positive value at late time (i.e. at present epoch).  
Recent cosmological observations (Perlmutter et al. 1998, 1999; Riess et al. 1998, 2004; Tonry et al. 2003) suggest the 
existence of a positive cosmological constant $\Lambda$ with the magnitude $\Lambda(G\hbar/c^{3})\approx 10^{-123}$. 
These observations on magnitude and red-shift of type Ia supernova suggest that our universe may be an accelerating 
one with induced cosmological density through the cosmological $\Lambda$-term. Thus, the nature of $\Lambda$ in our 
derived DE model is supported by recent observations. \\

Figure $4$ is the plot of deceleration parameter $q$ versus time $t$. From Figure $4$, it is observed that $q$ 
decreases very rapidly and reaches values $-1$ then after it remains constant $-1$ (like de Setter universe). 
From this figure we observe that the DE model, for $0 < n < 1.5$, evolves from the matter dominated era to quintessence 
era and ultimately approaches to cosmological constant era where as for $n \geq 1.5$, the universe evolves from   
quintessence to cosmological constant era. It is worth mentioned here that for $n < 1.5$, transition of the universe 
takes place from the early decelerating phase to the recent accelerating phase where as for $n \geq 1.5$, the expansion 
of the universe is always accelerating. \\

From these analysis we conclude that it is the choice of scale factor which makes the model inflationary at the early stages 
of the universe and radiation/matter dominance phase before the D.E. era. From Eq. (\ref{eq29}), we observe that when $t \to 0$, 
the expansion scalar $\theta$ becomes infinity which indicates the inflationary scenario. Also from Fig. $4$, we observe that 
before $t \approx 1$, $q > 0$ and this indicates radiation/matter dominance era of the universe. However, after $t \approx 1$, 
$q < 0$ which indicates the DE dominated era. The solution in our model does not blow up at any given epoch for the choice 
of the {\it ansatz} (\ref{eq20}). Hence our derived model is physically acceptable. \\

{\it The Cosmic Microwave Background} (CMB) is also considered to be a major experimental evidence which support the 
present models of the observed universe and from this CMB observations several scientists found the signature of anisotropy.
Based on these studies and observations, one may not preclude the possibility that our universe is anisotropic. We have 
already discussed this scenario in Introduction.
\begin{figure}[ht]
\centering
\includegraphics[width=10cm,height=10cm,angle=0]{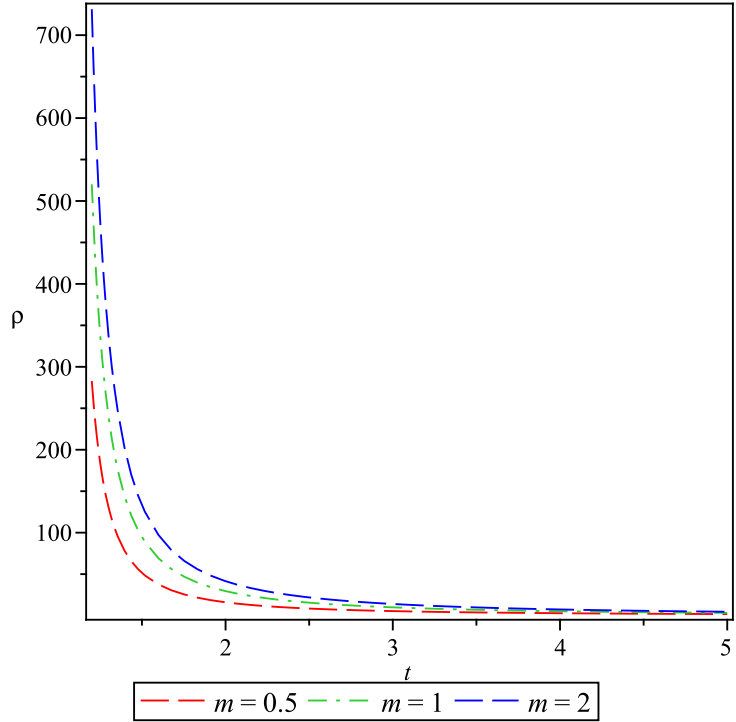} \\
\caption{The plot of energy density$\rho$ versus $t$. 
Here $\ell_{0} = 0.1$ }.
\end{figure}
\begin{figure}[ht]
\centering
\includegraphics[width=10cm,height=10cm,angle=0]{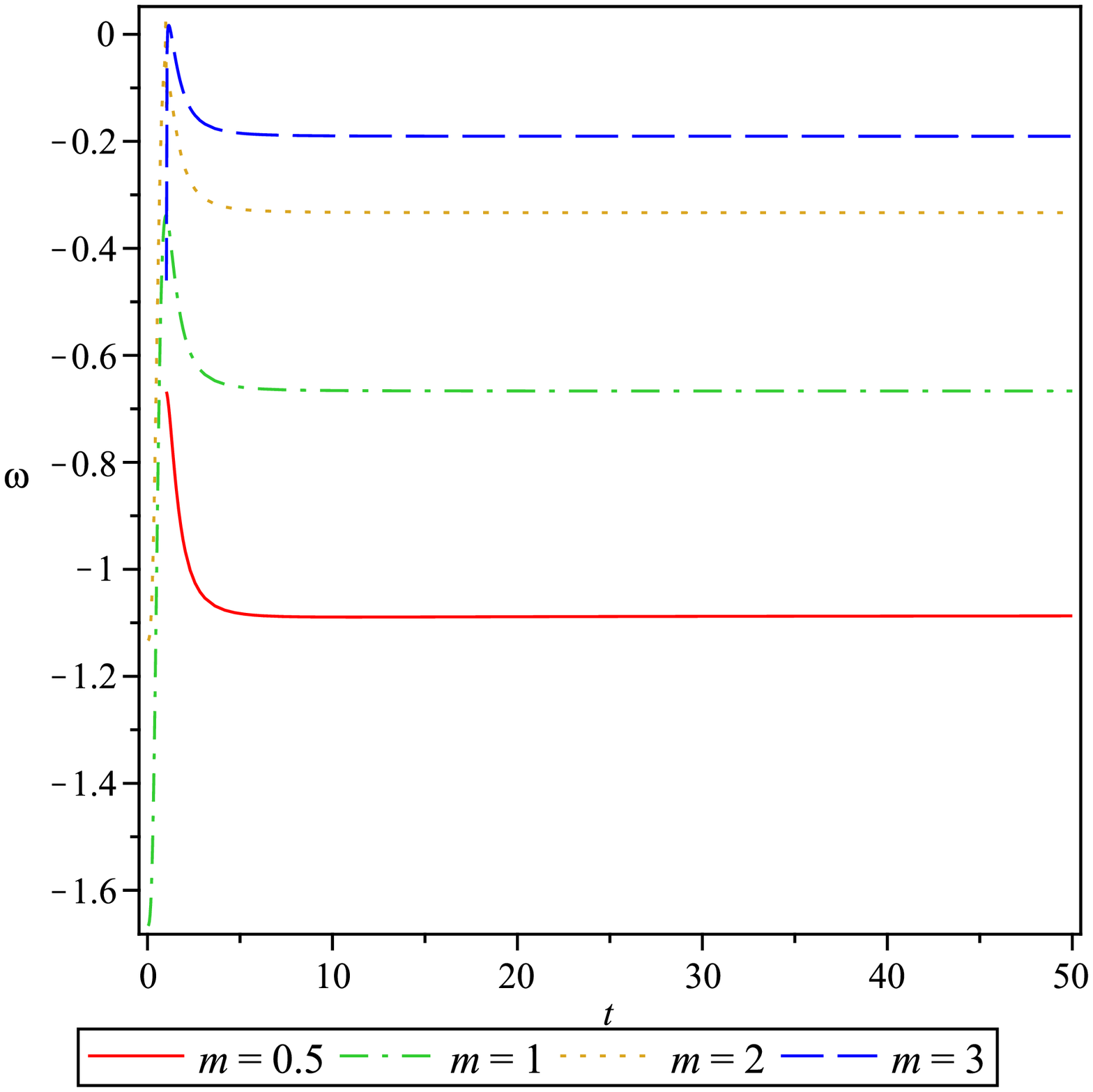} \\
\caption{The plot of EoS parameter $\omega$ versus $t$. 
Here $\ell_{0} = 0.1$}.
\end{figure}
\begin{figure}[ht]
\centering
\includegraphics[width=10cm,height=10cm,angle=0]{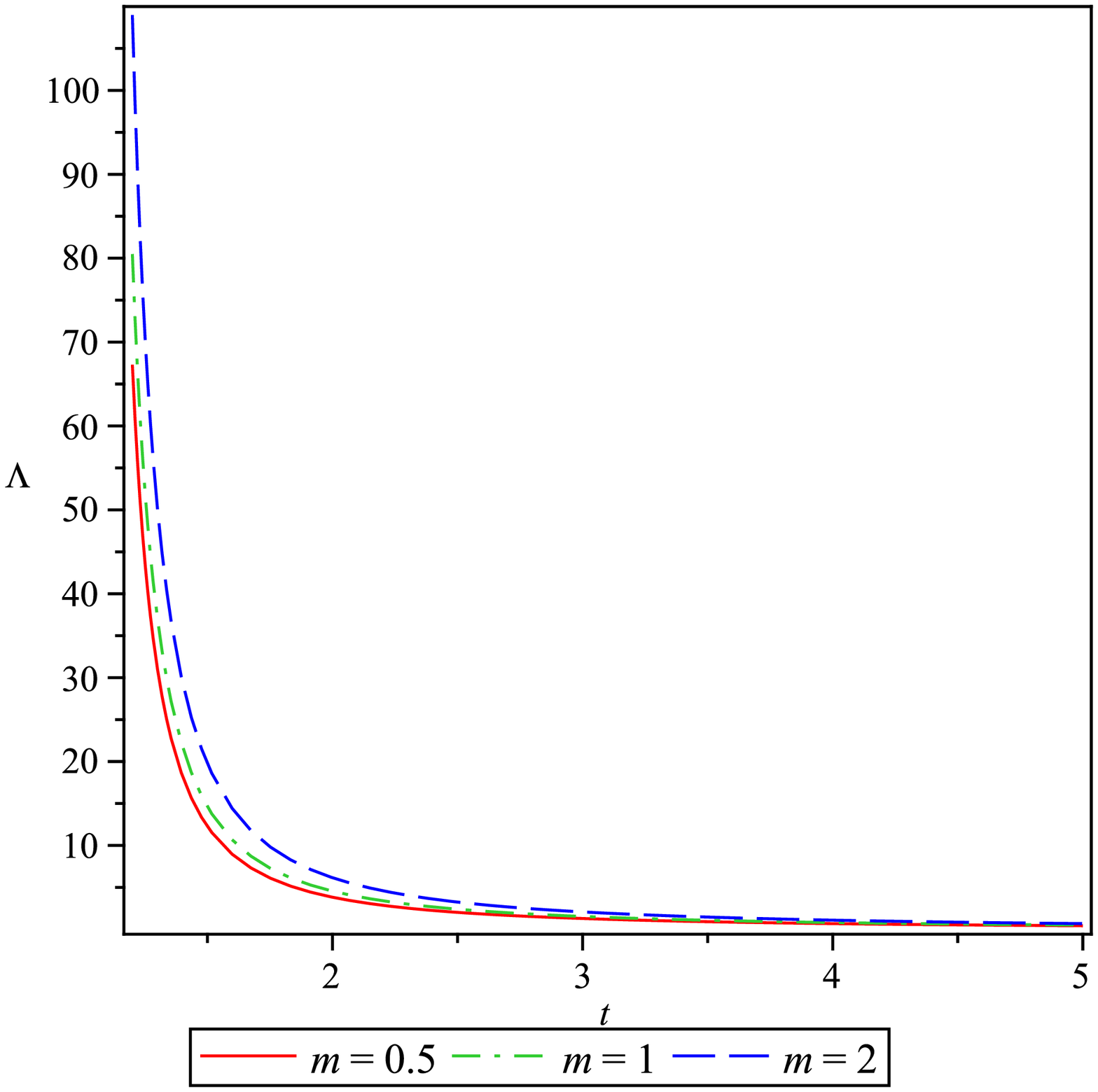} \\
\caption{The plot of cosmological constant $\Lambda$ versus $t$.
Here $\ell_{0} = 0.1$}.
\end{figure}
\begin{figure}[ht]
\centering
\includegraphics[width=10cm,height=10cm,angle=0]{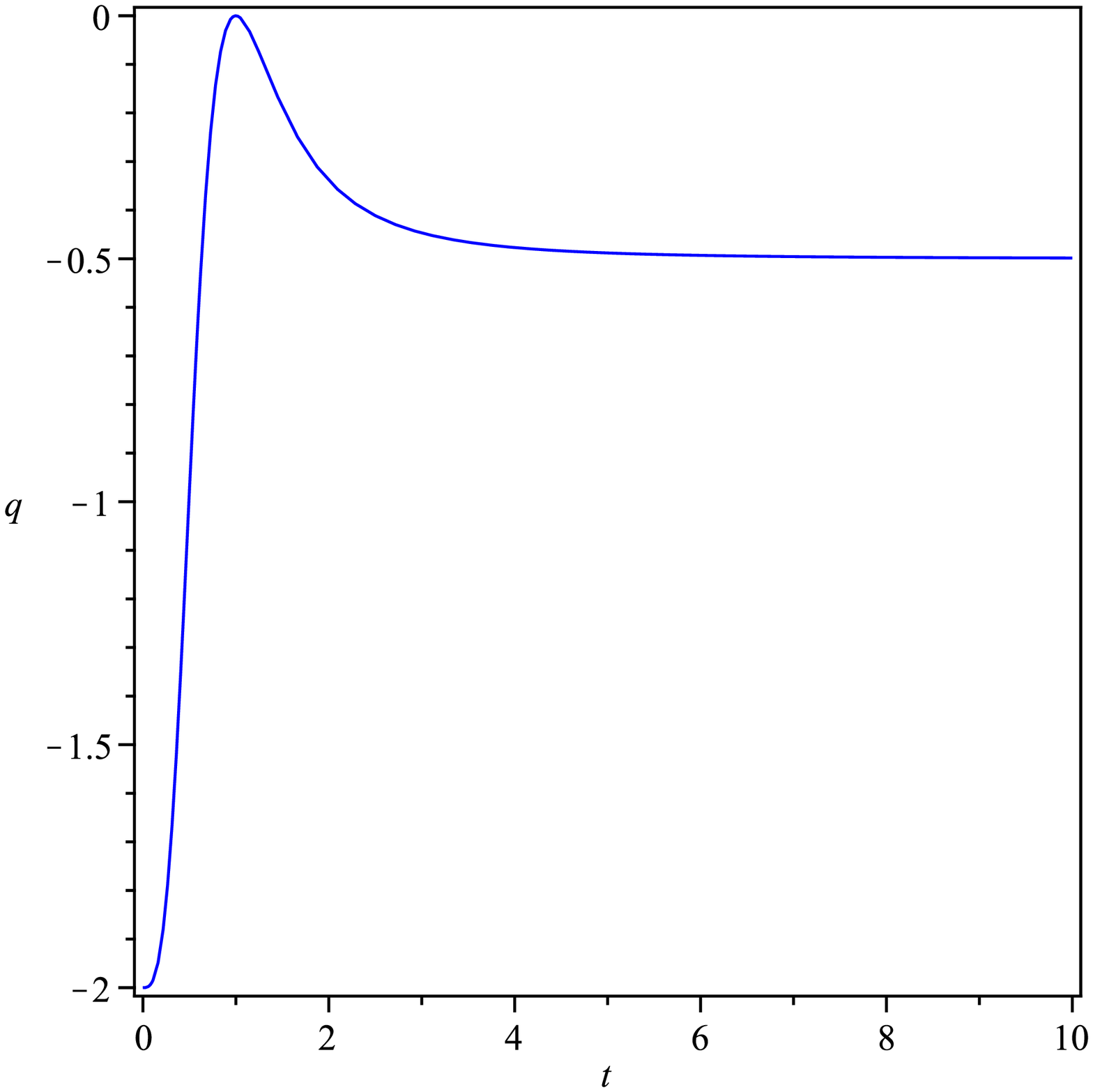} \\
\caption{The plot of deceleration parameter $q$ versus $t$}. 
\end{figure}
\section{STABILITY OF CORRESPONDING SOLUTIONS}
A rigorous analysis on the stability of the corresponding solutions can be done by invoking a perturbative approach.
Perturbations of the fields of a gravitational system against the background evolutionary solution should be checked to 
ensure the stability of the exact or approximated background solution (Chen and Kao 2001). Now we will study the stability 
of the background solution with respect to perturbations of the metric. Perturbations will be considered for all three 
expansion factors $a_{i}$ via
\begin{equation}
\label{eq43}
a_{i}\rightarrow a_{Bi}+\delta a_{i}=a_{Bi}(1+\delta b_{i})
\end{equation}
We will focus on the variables $\delta b_{i}$ instead of $\delta a_{i}$ from now on for convenience. Therefore, the 
perturbations of the volume scale factor $V_{B} = \Pi_{i=1}^{3}a_{i}$, directional Hubble factors $\theta_{i} = 
\frac{\dot{a_{i}}}{a_{i}}$ and the mean Hubble factor $\theta = \sum_{i=3}^{3}\frac{\theta_{i}}{3} = 
\frac{\dot{V}}{3V}$ can be shown to be
\begin{equation}
\label{eq44}
V\rightarrow V_{B} + V_{B}\sum_{i}\delta b_{i}, ~~~~ \theta_{i}\rightarrow \theta_{Bi}+\sum_{i}\delta 
b_{i}, ~~~~\theta\rightarrow \theta_{B}+\frac{1}{3}\sum_{i}\delta b_{i}
\end{equation}
One can show that the metric perturbations $\delta b_{i}$, to the linear order in $\delta b_{i}$, obey the following equations
\begin{equation}
\label{eq45}
\sum_{i}\delta\ddot{b_{i}}+2\sum\theta_{Bi}\delta\dot{b_{i}}=0
\end{equation}
\begin{equation}
\label{eq46}
\delta\ddot{b_{i}}+\frac{\dot{V}_{B}}{V_{B}}\delta \dot{b_{i}}+\sum_{j}\delta\dot{b_{j}}\theta_{Bi}=0
\end{equation}
\begin{equation}
\label{eq47}
\sum\delta\dot{b_{i}}=0.
\end{equation}
From above three equations, we can easily find
\begin{equation}
\label{eq48}
\delta \ddot{b_{i}}+\frac{\dot{V}_{B}}{V_{B}}\delta \dot{b_{i}}=0,
\end{equation}
where $V_{B}$ is the background volume scale factor. In our case, $V_{B}$ is given by
\begin{equation}
\label{eq49}
V_{B}=t^{\frac{3}{2}n}e^{\frac{3}{2}t}
\end{equation}
using above equation in equation (6) and after integration we get
\begin{equation}
\label{eq50}
\delta b_{i} = c_{i} t^{-\frac{3}{4}n}e^{-\frac{3}{4}t}\mbox{WittakerM}\left(-\frac{3}{4}n,-\frac{3}{4}n + 
\frac{1}{2},\frac{3}{2}t\right),
\end{equation} 
where $c_{i}$ is an integration constant. Therefore, the ``actual`` fluctuations for each expansion factor 
$\delta a_{i} = a_{Bi}\delta b_{i}$
is given by
\begin{equation}
\label{eq51}
\delta a_{i}\rightarrow c_{i} t^{-\frac{n}{4}}e^{-\frac{t}{4}}\mbox{WittakerM}\left(-\frac{3}{4}n,-\frac{3}{4}n + 
\frac{1}{2},\frac{3}{2}t\right).
\end{equation} 
From above equation we see that for $n> > 1$, $\delta a_{i}$ approaches zero. Consequently, the background solution is 
stable against the perturbation of the graviton field. 

\section{OTHER DARK ENERGY MODEL}
Now we take the following {\it ansatz} for the scale factor, where the increase in terms of time evolution is 
\begin{equation}
\label{eq52}
a(t) = -\frac{1}{t} + t^{2}.
\end{equation}
By the above choice of scale factor yields a time dependent deceleration parameter and the corresponding solutions are 
stable. The motivations for selecting such type of scale factors for finding solutions are already described in Sect. $3$. 
We define the deceleration 
parameter $q$ as usual,
\begin{equation}
\label{eq53}
q = - \frac{\ddot{a}a}{\dot{a}^{2}} = - \frac{\ddot{a}}{aH^{2}}.
\end{equation}
Using (\ref{eq52}) into (\ref{eq53}), we find
\begin{equation}
\label{eq54}
q = -2\left(\frac{t^{3} - 1}{2t^{3} + 1}\right)^{2}.
\end{equation}
Using equations (\ref{eq16}), (\ref{eq24}) and (\ref{eq52}) in (\ref{eq11}), we obtain the expressions for 
metric functions as follows 
\begin{equation}
\label{eq55}
B(t) = \ell_{4}\left(-\frac{1}{t} + t^{2}\right)^{\frac{3}{(m + 2)}},
\end{equation}
\begin{equation}
\label{eq56}
C(t) = \ell_{5}\left(-\frac{1}{t} + t^{2}\right)^{\frac{3}{(m + 2)}},
\end{equation}
\begin{equation}
\label{eq57}
A(t) =  \ell_{6}\left(-\frac{1}{t} + t^{2}\right)^{\frac{3m}{(m + 2)}},
\end{equation}
where, $\ell_{4} = \l^{-\frac{1}{(m + 2)}}, ~ \ell_{5} = \ell \ell_{4}, ~ \ell_{6} = \ell_{4}^{m}$ and 
$\l$ is an integrating constant. \\

Hence the model (\ref{eq1}) reduces to
\[
ds^{2} = - dt^{2} + \ell^{2}_{6}\left(-\frac{1}{t} + t^{2}\right)^{\frac{6m}{(m + 2)}} dx^{2} +  
\ell^{2}_{4}\left(-\frac{1}{t} + t^{2}\right)^{\frac{6}{(m + 2)}} dy^{2}
\]
\begin{equation}
\label{eq58}
+ \; \ell^{2}_{5}\left(-\frac{1}{t} + t^{2}\right)^{\frac{6}{(m + 2)}}dz^{2}.
\end{equation}
The expressions for the Hubble parameter ($H$), scalar of expansion ($\theta$), shear scalar ($\sigma$), spatial volume 
($V$) and the average anisotropy parameter ($A_{m}$) for the model (\ref{eq58}) are given by
\begin{equation}
\label{eq59}
\theta = 3H = \frac{3}{t}\left(\frac{2t^{3} + 1}{t^{3} - 1}\right),
\end{equation}
\begin{equation}
\label{eq60}
\sigma^{2} = 3\left[\left(\frac{m - 1}{m + 2}\right)\frac{(2t^{3} + 1)}{(t^{3} - 1)t}\right]^{2}.
\end{equation}
\begin{equation}
\label{eq61}
V = \left(-\frac{1}{t} + t^{2}\right)^{3},
\end{equation}
\begin{equation}
\label{eq62}
A_{m} =  2\left(\frac{m - 1}{m + 2}\right)^{2}.
\end{equation}
From Eq. (\ref{eq59}), we observe that when $t \to 0$, $\theta \to \infty$ and this indicates the inflationary scenario at 
early stages of the universe. Since $\frac{\sigma^{2}}{\theta^{2}} \ne 0 $ for all values of $m$ except for $m = 1$, hence 
the model is anisotropic except for $m = 1$. The the dynamics of the mean anisotropy parameter depends on the value of $m$. 
The mean anisotropic parameter is constant. We observed that when $m = -2$, $A_{m} \to \infty$ and for $m = 1$, 
$A_{m} = 0$. Thus, the observed isotropy of the universe can be achieved in phantom model (see, Figure $6$). \\

The energy density of the fluid can be find by using Eqs. (\ref{eq55}) \& (\ref{eq57}) in (\ref{eq19})
\begin{equation}
\label{eq63}
\rho = \frac{9(2m + 1)}{(m + 2)^{2}}\frac{(2t^{3} + 1)^{2}}{(t^{3} - 1)^{2}t^{2}} - 
\ell_{0}\left(-\frac{1}{t} + t^{2}\right)^{-\frac{6m}{(m + 2)}}.
\end{equation}
where $\l_{0} = \frac{1}{\ell_{6}^{2}}$. Using Eqs. (\ref{eq55}), (\ref{eq57}) and (\ref{eq63}) in (\ref{eq17}), 
the EoS parameter $\omega$ is obtained as
\begin{equation}
\label{eq64}
\omega = \frac{\frac{27}{(m + 2)^{2}}\frac{(2t^{3} + 1)^{2}}{(t^{3} - 1)^{2}t^{2}} - \frac{6}{(m + 2)}\frac{(2t^{6} + 
8t^{3} - 1)}{(t^{3} - 1)^{2}t^{2}} + \ell_{0}\left(-\frac{1}{t} + t^{2}\right)^{-\frac{6m}{(m + 2)}}}
{\ell_{0}\left(-\frac{1}{t} + t^{2}\right)^{-\frac{6m}{(m + 2)}} + \frac{9(2m + 1)}{(m + 2)^{2}}\frac{(2t^{3} + 1)^{2}}
{(t^{3} - 1)^{2}t^{2}}}.
\end{equation}
Using Eqs. (\ref{eq55}), (\ref{eq57}), (\ref{eq63}) and (\ref{eq64}) in (\ref{eq18}), the 
skewness parameters $\delta$ (or $\gamma$) (i.e. deviations from $\omega$ along y-axis and z-axis) 
are computed as
\begin{equation}
\label{eq65}
\delta = \gamma =  \frac{6\left(\frac{m - 1}{m + 2}\right)\frac{(5t^{6} + 2t^{3} + 2)}{(t^{3} -)^{2}t^{2}} - 
2\ell_{0}\left(-\frac{1}{t} + t^{2}\right)^{-\frac{6m}{(m + 2)}}}{\ell_{0}\left(-\frac{1}{t} + 
t^{2}\right)^{-\frac{6m}{(m + 2)}} - \frac{9(2m + 1)}{(m + 2)^{2}}\frac{(2t^{3} + 1)^{2}}{(t^{3} - 1)^{2}t^{2}}}.
\end{equation}
So, if the present work is compared with experimental results (Knop et al. 2003; Tegmark et al. 2004b; Hinshaw et al. 
2009; Komatsu et al. 2009), then one can conclude that the limit of $\omega$ provided by Eq. (\ref{eq64}) may 
accommodated with the acceptable range of EoS parameter. Also it is observed that at $t = t_{c}$, $\omega$ vanishes, 
where $t_{c}$ is a critical time given by the following relation
\begin{equation}
\label{eq66}\frac{27}{(m + 2)^{2}}\frac{(2t_{c}^{3} + 1)^{2}}{(t_{c}^{3} - 1)^{2}t_{c}^{2}} - \frac{6}{(m + 2)}
\frac{(2t_{c}^{6} + 8t_{c}^{3} - 1)}{(t_{c}^{3} - 1)^{2}t_{c}^{2}} + \ell_{0}\left(-\frac{1}{t_{c}} + t_{c}^{2}
\right)^{-\frac{6m}{(m + 2)}} = 0.
\end{equation}
Thus, for this particular time, our model represents a dusty universe. We also note that the earlier real matter at 
$t \leq t_{c}$, where $\omega \geq 0$ later on at $t > t_{c}$, where $\omega < 0$ converted to the dark energy 
dominated phase of universe. \\

From Eq. (\ref{eq63}), we note that energy density of the fluid $\rho(t)$ is a decreasing function of time and 
$\rho \geq 0$ when
\begin{equation}
\label{eq67}  \frac{(2t^{3} + 1)^{2}}{(t^{3} - 1)^{2}t^{2}}\left(-\frac{1}{t} + t^{2}\right)^{\frac{6m}{(m + 2)}}
\geq \frac{\ell_{0}(m + 2)^{2}}{9(2m + 1)}.
\end{equation}  
Figure $5$ is the plot of energy density of the fluid ($\rho$) versus time $t$. Here we observe that $\rho$ is 
a positive decreasing function of time and it approaches to zero as $t \to \infty$. \\

Figure $6$ depicts the variation of EoS parameter ($\omega$) versus cosmic time ($t$) in evolution of the universe, 
as a representative case with appropriate choice of constants of integration and other physical parameters using 
reasonably well known situations (parameters are given in Figure caption). From Figure $6$, we observe as follows: \\

(i) for $m \leq 0.5$, the evolution of the universe starts from quintessence era ($\omega > -1$) and approaches to 
phantom region ($\omega < -1$). \\

(ii) for $1 \leq m <2$, the universe evolves from phantom region ($\omega < -1$), then crosses PDL and ultimately 
approaches to quintessence region ($\omega > -1$). \\

(iii) for $2 \leq m \leq 3$, the evolution of the universe commence from phantom region ($\omega < -1$), then crosses 
PDL and then skip over to non-dark region. \\

(iv) for $3 \leq m$, the evolution of the universe begins from quintessence era ($\omega > -1$) and ultimately pass over 
to non-dark region. \\

(v) for $m = 1$, we get $\omega \approxeq -0.65$ which is consistent with SNe Ia data $-1.67 < \omega < -0.62$ 
(Knop et al. 2003). \\

(vi) for $m = 0.5$, we get $\omega \approxeq -1.1$ which is reproducible with current observational realm (Knop et al. 2003; 
Tegmark et al. 2004b; Hinshaw et al. 2009; Komatsu et al. 2009). \\

Using Eqs. (\ref{eq59}) and (\ref{eq63}) in (\ref{eq40}), the cosmological constant is obtained as 
\begin{equation}
\label{eq68} \Lambda = \frac{3(4m^{2} + 10m + 13)}{(m + 2)^{2}}\frac{(2t^{3} + 1)^{2}}{(t^{3} - 1)^{2}t^{2}} + 
\ell_{0}\left(-\frac{1}{t} + t^{2}\right)^{\frac{-6m}{(m + 2)}}.
\end{equation}
From Eq. (\ref{eq68}), we observe that $\Lambda$ is a decreasing function of time and it is always positive when
\begin{equation}
\label{eq69} \frac{(2t^{3} + 1)^{2}}{(t^{3} - 1)^{2}t^{2}}\left(-\frac{1}{t} + t^{2}\right)^{\frac{6m}{(m + 2)}}
> - \frac{\ell_{0}(m + 2)^{2}}{3(4m^{2} + 10m + 13)}. 
\end{equation}
Figure $7$ is the plot of cosmological constant $\Lambda$ versus time $t$. It is observed that in all cases 
cosmological parameter is decreasing function of time and it approaches a small positive value at late time 
(i.e. at present epoch). Thus, the nature of $\Lambda$ in this derived DE model is also in good agreement 
with recent observations (Perlmutter et al. 1998, 1999; Riess et al. 1998, 2004; Tonry et al. 2003). \\

Figure $8$ is the plot of deceleration parameter $q$ versus time $t$. From the figure we observe that the expansion 
of the universe starts from accelerating phase and the rate of expansion decreases with time and it stops and again 
starts accelerating to approach $-0.5$ which is very close to the value ($\approx -0.7$) predicted by the observations 
(Riess et al. 2004; Virey et al. 2005). \\

A convenient method to describe models close to $\Lambda$ CDM is based on the cosmic jerk parameter $j$, a
dimensionless third derivative of the scale factor with respect to the cosmic time (Chiba and Nakamura 1998; 
Sahni 2002; Blandford et al. 2004; Visser 2004, 2005). A deceleration-to-acceleration transition 
occurs for models with a positive value of $j_{0}$ and negative $q_{0}$. Flat $\Lambda$ CDM models have a constant 
jerk $j = 1$. The jerk parameter in this case is obtained as 
In this case, we obtain the jerk parameter as
\begin{equation}
\label{eq70} j(t) = \frac{2t^{5} + 2t^{4} - 2t^{2} - t -2}{(t + 1)(1 + t^{2})}.
\end{equation}
This value is consistent with observational value $j\simeq2.16$ obtained from the combination of three kinematical 
data sets: the gold sample of type Ia supernovae (Riess et al. 2004), the SNIa data from the SNLS project (Astier 
et al. 2006), and the X-ray galaxy cluster distance measurements (Rapetti et al. 2007) for $t = 1.50$.  
\section{STABILITY OF CORRESPONDING SOLUTIONS}
The method to study the stability of the background solution with respect to perturbations of the metric is already 
given in Sect. $5$. From Eqs. (\ref{eq45})$-$(\ref{eq47}), we can easily derive 
\begin{equation}
\label{eq71}
\delta \ddot{b_{i}}+\frac{\dot{V}_{B}}{V_{B}}\delta \dot{b_{i}}=0,
\end{equation}
 where $V_{B}$ is the background volume scale factor. In our case, $V_{B}$ is given by 
\begin{equation}
\label{eq72} V_{B} = t^{6}.
\end{equation}
Using above equation in Eq. (\ref{eq71}) and after integration we get
\begin{equation}
\label{eq73} \delta b_{i} = c_{i}t^{-5},
\end{equation}
where $c_{i}$ is an integration constant. Therefore, the ``actual`` fluctuations for each expansion factor $\delta a_{i} = 
a_{Bi}\delta b_{i}$ is given by
\begin{equation}
\label{eq74} \delta a_{i}\rightarrow c_{i} t^{-3},
\end{equation}
where $a_{Bi}\rightarrow t^{2}$. From above equation it is obvious that $\delta a_{i}$ approaches zero as $t \to \infty$. 
Consequently, the background solution is stable against the perturbation of the graviton field.

\section{CONCLUDING REMARKS}
A new class of anisotropic B-$VI_{0}$ DE models with variable EoS parameter $\omega$ has been investigated by 
using time dependent deceleration parameter. In literature it is plebeian to practice a constant deceleration 
parameter. Now for a Universe which was deceleration in past and accelerating at present epoch, the DP must 
show signature flipping as already discussed in Section $2$. Therefore our consideration of DP to be variable 
is physically justified. \\

The main features of the models are as follows: \\

\begin{itemize}
\item DE models present the dynamics of EoS parameter $\omega$ provided by Eqs. (\ref{eq35}) and (\ref{eq55}) 
whose range are in good agreement with the acceptable range by the recent observations (Knop et al. 2003; Tegmark 
et al. 2004b; Hinshaw et al. 2009; Komatsu et al. 2009). 

\item It can be easily seen that in both DE models, the mean anisotropic parameter vanishes at $m = 1$. Thus, our 
both anisotropic models approach to isotropy at $m = 1$. 

\item We obtain cosmological constant dominated universe, quintessence and phantom fluid dominated universe (Chevallier 
and Polarski  2001), representing the different phases of the universe through-out the evolving process for different cosmic 
times. These fits suggest that $\omega > -1$ for a long (quintessence-like) period in the past, and at the same time they 
suggest that the universe has just entered a phantom phase $\omega < -1$ near our present. 

\item Unlike Robertson-Walker (RW) metric Bianchi type metrics can admit a DE that wields an anisotropic EoS parameter 
according to the characteristics. Therefore, one can not rule out the possibility of anisotropic nature of DE  in the 
frame-work of B-$VI_{0}$ space-time. 

\item In first case, the observed isotropy of the universe can be achieved in cosmological constant model (see, Figure $2$) 
whereas in second case, the observed isotropy of the universe can be achieved in phantom model (see, Figure $6$). Thus, 
Bianchi type-$VI_{0}$ models which remain anisotropic are of preferably academical interest. 

\item Our DE models are of great importance in the sense that the nature of decaying vacuum energy density $\Lambda(t)$ is
supported by recent cosmological observations (Perlmutter et al. 1998, 1999; Riess et al. 1998, 2004; Tonry et al. 2003). 

\item Though there are many suspects (candidates) such as cosmological constant, vacuum energy, scalar field, brane world, 
cosmological nuclear-energy, etc. as reported in the vast literature for DE, the proposed models in this paper favour 
EoS parameter as a possible suspect for the DE. 

\item The cosmic jerk parameter in our derived models is also found to be in good agreement with the recent data of astrophysical 
observations namely the gold sample of type Ia supernovae (Riess et al. 2004), the SNIa data from the SNLS project 
(Astier et al. 2006), and the X-ray galaxy cluster distance measurements (Rapetti et al. 2007). 

\item For different choice of $n$, we can generate a class of DE models in Bianchi type-$VI_{0}$ space-time. It is observed 
that such DE models are also in good harmony with current observations. Our study is continued and we shall generate some other 
interesting physically viable models for other values on $n$.

\item Our corresponding solutions have inflationary scenario at the early stages of the universe and also radiation/matter 
era before DE era. 

\item Our corresponding solutions are physically acceptable and the solutions are stable. \\
\end{itemize}
Thus, the solutions demonstrated in this paper may be useful for better understanding of the characteristic of anisotropic DE 
in the evolution of the universe within the framework of Bianchi type-$VI_{0}$ space-time.
\section*{Acknowledgments} 
The author would like to thank IUCAA, Pune, India for providing facility and support for a visit under associateship program 
where part of this work was carried out. The financial support (Project No. C.S.T./D-1536) in part by the State Council of 
Science and Technology, Uttar Pradesh (U. P.), India is gratefully acknowledged. We also thank Hassan Amirhashchi, Farook 
Rahaman and P. K. Samal for helpful discussions. 

\end{document}